\documentclass[11pt,notitlepage,nofootinbib]{revtex4-2}
\usepackage[width=480pt,height=640pt]{geometry}

\usepackage[dvipdf,dvips,dvipdfmx]{graphicx}
\usepackage
	[breaklinks,linktoc=all,colorlinks=true,linkcolor=blue,citecolor=blue,urlcolor=blue]
	{hyperref}
	\renewcommand\thefootnote{\textcolor{blue}{\arabic{footnote}}}

\usepackage{xurl}
\usepackage{epsfig}
\usepackage{epstopdf}
\usepackage{float}
\usepackage{slashed}
\usepackage{cancel}
\usepackage{textcomp}
\usepackage{adjustbox}
\usepackage{enumitem}
\usepackage{bbm}
\usepackage{bbold}
\usepackage{indentfirst}
\usepackage{wrapfig}
\usepackage{comment}
\usepackage{soul,xcolor}
  \setstcolor{red}
\usepackage{pgfplots}
	\pgfplotsset{width=5cm,compat=1.9}
\usepackage{bm}
\usepackage{fancyhdr}
\usepackage{amsmath}
\usepackage{amssymb}
\usepackage{calc}
\usepackage{mathtools}
\usepackage{gauss}
\usepackage{physics}
\usepackage{gensymb}
\usepackage{esint}
\usepackage{tikz-feynman}
\usepackage{tensor}
\usepackage{xfrac}
	
\allowdisplaybreaks

\newcommand{\nnr}{\nonumber\\}

\def\lsim{\mathrel{\mathpalette\@versim<}}
\def\gsim{\mathrel{\mathpalette\@versim>}}

\newcommand{\dx}{\dd^4x\,}

\newcommand{\Mp}{M_\text{p}}

\newcommand{\met}{g_{\alpha\beta}}

\newcommand{\gdet}{\sqrt{-g}}

\newcommand{\Hab}{H_{\alpha\beta}}
\newcommand{\iHab}{H^{\alpha\beta}}
\newcommand{\Hs}[1]{H_{#1}{}^{#1}}

\newcommand{\ssz}[1]{\tilde{#1}}

\newcommand{\Lag}{\mathcal{L}}
\newcommand{\Act}{\mathcal{S}}
\newcommand{\Ham}{\mathcal{H}}

\def\bfx{{\boldsymbol x}}

\def\bfk{{\boldsymbol k}}

\def\Re{{\rm Re}\,}
\def\Im{{\rm Im}\,}

\def\sqr#1#2{{\vcenter{\hrule height.#2pt
      \hbox{\vrule width.#2pt height#1pt \kern#1pt
          \vrule width.#2pt}
      \hrule height.#2pt}}}

\begin{document}

\title{Primordial Gravitational Waves in Quadratic Gravity}
\maketitle
\vspace{-12pt}

\renewcommand{\thefootnote}{\fnsymbol{footnote}}

\begin{center}
{\large
Jisuke Kubo$^{1,2,}$\footnote{jikubo4@gmail.com} and Jeffrey Kuntz$^{1,}$\footnote{jkuntz@mpi-hd.mpg.de}} \\ \vspace{8pt}
$^1${\it 
Max-Planck-Institut f\"ur Kernphysik(MPIK), Saupfercheckweg 1, 69117 Heidelberg, Germany} \\ \vspace{4pt}
$^2${\it 
Department of Physics, University of Toyama, 3190 Gofuku, Toyama 930-8555, Japan} \\ \vspace{8pt}
(Dated: \today)
\\ \vspace{8pt}
\end{center}

\renewcommand{\thefootnote}{\arabic{footnote}}
\setcounter{footnote}{0}

\section*{Abstract}

Quadratic gravity is a fourth-order (in derivatives) theory that can serve as an attractive upgrade to the standard description of gravity provided by General Relativity, thanks to its renormalizability and its built-in description of primordial inflation. We bring quadratic gravity into a second-order form by introducing an auxiliary tensor field and we consider the primordial tensor fluctuations (gravitational waves) in the theory around a Friedmann-Lemaître-Robertson-Walker background. After a canonical quantization of the perturbations, we calculate the tensor power spectrum in quasi de Sitter spacetime. We find that the spectral index $n_t$ and the amplitude $A_t$ of the tensor power spectrum are both suppressed by the factor $(1 + 2 \bm H^2_*/m_\text{gh}^2)^{-1}$, where $\bm H_*$ is the Hubble rate at horizon exit and $m_\text{gh}$ is the mass of the spin-two ghost. This restores the slow-roll consistency condition familiar from single-field inflation models, where the tensor-to-scalar ratio $r$ is equal to $-8n_t$ in the lowest nontrivial order in the slow-roll approximation. We also discuss the well-known issue of the ghost problem in fourth-order theories and how it pertains to the results at hand.

\clearpage
\section{Introduction}

The primordial quantum fluctuations generated during cosmic inflation are known to be the origin of the observed anisotropies in the cosmic microwave background (CMB), as well as the seeds for structure formation in the early universe \cite{Mukhanov:1981xt,Hawking:1982cz,Starobinsky:1982ee,Guth:1982ec,Bardeen:1983qw} (see also \cite{Kodama:1984ziu,Copeland:1993jj,Bartolo:2004if,Weinberg2008,Baumann:2009ds} and references therein). These fluctuations are quantified in terms of the $n$-point correlation functions (see e.g \cite{Copeland:1993jj,Bartolo:2004if,Weinberg2008,Baumann:2009ds}). The two-point correlation function (power spectrum) of scalar curvature perturbations is of particular importance as it is parameterized by the \textit{observable} spectral index $n_s$ and amplitude $A_s$ \cite{Lyth:1984yz,Lyth:1984gv,Mukhanov:1985rz,Sasaki:1986hm,Stewart:1993bc}. Cosmic inflation also predicts the existence of tensor fluctuations (primordial gravitational waves) whose power spectrum is parameterized, as in the case of the scalar curvature perturbations, by the index $n_t$ and the amplitude $A_t$ \cite{Starobinsky:1979ty,Rubakov:1982df,Fabbri:1983us,Abbott:1984fp,Copeland:1993jj}. Interestingly, tensor fluctuations have so far not been observed, a fact which strongly constrains the tensor-to-scalar ratio $r=A_t/A_s < 0.036$ at $95$ C.L. \cite{BICEP:2021}, and hints that the Starobinsky model \cite{Starobinsky:1980te} seems to be the best candidate theory for inflation \cite{Planck:2018jri,BICEP:2021}. It will in fact be possible to put even stronger constraints on the parameters mentioned above in the near future. The space-based experiment LiteBird \cite{LiteBIRD:2022cnt} and the earth-based experiment CMB-S4 \cite{CMB-S4:2020lpa}, for instance, will be sensitive down to the order of the prediction of the Starobinsky model, $r \sim O(10^{-3})$, while another proposed space-based experiment, PICO, will attempt to measure $r$ at the level of $O(10^{-4})$ \cite{NASAPICO:2019thw} (see also \cite{Achucarro:2022qrl} for other future experiments).

The essential term in the Lagrangian of the Starobinsky model is $R^2$ (the square of the Ricci curvature scalar), though as a quantum field theory (QFT), this model is not renormalizable. There does however exist a renormalizable theory of gravity, dubbed quadratic gravity (QG) \cite{Stelle:1976gc}, that contains all independent curvature operators of dimension at most four in its Lagrangian. Moreover, given that the $R^2$ is of dimension four, it belongs to this class of operators and one may naturally embed the Starobinsky model into the framework of renormalizable QG so as to include all possible renormalizable operators, including the squared Weyl tensor $C_{\alpha\beta\gamma\delta}^2$, while maintaining the crucial inflationary features \cite{Clunan:2009er,Deruelle:2010kf,Deruelle:2012xv,Myung:2014jha,Myung:2015vya,Baumann:2015xxa,Ivanov:2016hcm,Salvio:2017xul,Ghilencea:2019rqj,Edery2019,Anselmi:2020lpp,Salvio:2020axm,Kubo:2022dlx} (see also \cite{DeFelice:2023psw} for a contrary finding).

Though these features make QG an appealing theory, there are caveats. The squared Weyl tensor that appears in the QG Lagrangian is an indispensable ingredient to ensure renormalizability of the theory as it in particular leads to a softening of the high energy behavior \cite{Stelle:1976gc}. Unfortunately however, due to the fact that this Weyl term contains four derivatives of the metric, QG predicts the existence of a massive spin-2 indefinite-metric ghost that endangers the physical unitarity of the theory \cite{Stelle:1977ry}. There have been many attempts to solve or get around this problem (see e.g.\ \cite{Lee:1969fy,Lee:1969zze,Lee:1970iw,Boulware:1983td,Mannheim:2011ds,Salvio:2018crh,Anselmi:2018tmf,Donoghue:2019fcb,Kuntz2024}), though these usually require one to step outside the framework of conventional QFT \cite{Kubo:2023lpz,Kubo:2024ysu}.

Given that the tensor-to-scalar $r$ ratio is one of the few observables that we can actually measure which might give us information about the nature of gravity beyond the usual Einstein picture, it is important to understand exactly how the tensor-to-scalar ratio predicted by QG differs from that of the bare $R^2$ Starobinsky model. The crucial differences stem from the fact that the tensor fluctuations in QG satisfy a fourth-order equation of motion due to the presence of the squared Weyl term in the Lagrangian. Because of this fact, the amplitude $A_t$ of the power spectrum is thought to pick up a correction to $r$ \cite{Deruelle:2012xv,Myung:2014jha,Myung:2015vya,Salvio:2017xul,Anselmi:2020lpp}:
\begin{align} \label{correction-to-r}
r \to r (1+2H_*^2/m_\text{gh}^2)^{-1} \,,
\end{align}
where $H_*$ is the Hubble parameter at the horizon exit. Using the Planck bound $H_*<5.7\times10^{-5} M_\text{Pl}$ \cite{Planck:2018jri}, one finds that the correction $2H_*^2/m_\text{gh}^2$ can become of $O(1)$ for $m_\text{gh}\lesssim10^{-4}M_\text{Pl}$. This implies that, if the tensor-to-scalar ratio $r$ were corrected as in (\ref{correction-to-r}) without changing the spectral index $n_t$, then the slow-roll consistency condition in single-filed inflation (to first order in the slow-roll approximation) \cite{Liddle:1992wi,Turner:1993xz,Copeland:1993jj} (see also \cite{Lidsey:1995np} and references therein),
\begin{align} \label{consistency}
r = -8 n_t \,,
\end{align}
would be violated. However, in \cite{Anselmi:2020lpp} the tensor power spectrum in QG is computed by implementing the fakeon prescription \cite{Anselmi:2018kgz,Anselmi:2018tmf} and it is found that the consistency condition (\ref{consistency}) is in fact satisfied.

The intention of the present paper is to see whether the consistency condition (\ref{consistency}) is satisfied within the framework of {\it conventional} QFT. To this end, we reformulate QG, which is fourth-order in derivatives in its bare form, into a second-order form and consider tensor perturbations around a Friedmann-Lemaître-Robertson-Walker (FLRW) background, which is done in Section \ref{sec:tenspurt}. We then perform a canonical quantization in Section \ref{sec:canquant}, where we find a (quasi) canonical transformation that diagonalizes the coupled canonical system in pure de Sitter spacetime. This enables us to compute the tensor power spectrum exactly, confirming the known results of \cite{Deruelle:2012xv,Myung:2014jha,Myung:2015vya,Salvio:2017xul,Anselmi:2020lpp}. In Section \ref{sec:quasi-dS} we extend our analysis to quasi de Sitter spacetime and find an approximate solution to a system of coupled DEs in the superhorizon that yields the tensor spectral index $n_t$ and confirms the consistency condition (\ref{consistency}). Despite this fact, we predict that the result obtained in our conventional, canonical method will be different from that obtained by using the fakeon prescription \cite{Anselmi:2020lpp} at the same level of precision, at higher orders in the slow-roll approximation. In Section \ref{sec:gprob}, we discuss the spin-two ghost problem in quadratic gravity and how it relates to inflationary models based on it. Finally, we then close with concluding remarks.

\section{Tensor perturbations in FLRW spacetime} \label{sec:tenspurt}

We begin with the complete action for quadratic gravity in the Einstein frame
\begin{align}
\Act_\text{QG} = \int\dx\gdet\bigg[\frac{\Mp^2}{2}R - \frac{\alpha^2}{2}C_{\alpha\beta\gamma\delta}C^{\alpha\beta\gamma\delta} + \beta^2R^2 + \Lag_\text{matter}[g]\bigg] \,,
\end{align}
which includes an arbitrary matter Lagrangian, an Einstein-Hilbert term, a Starobinsky $R^2$ term with dimensionless coupling $\beta$, and a square of the Weyl tensor with coupling $\alpha$ which, as mentioned in the Introduction, is crucial for the renormalizability of the theory. It is convenient to rewrite this last term using the relation $C_{\alpha\beta\gamma\delta}C^{\alpha\beta\gamma\delta} = 2R_{\alpha\beta}R^{\alpha\beta} - \frac{2}{3}R^2 + \mathcal{G}$ where $\mathcal{G}$ is a total derivative (the Gauss-Bonnet invariant) that may be neglected for our purposes. Additionally, since we will focus only on tensor fluctuations of the metric in what follows, we may also set $\beta=0$ since the Starobinsky term contributes only additional scalar fluctuations to the usual tensor fluctuations present in pure GR. With this, we are left with the action  
\begin{align} \label{SQG}
\Act = \int\dx\gdet\bigg[\frac{\Mp^2}{2}R - \alpha^2\bigg(R_{\alpha\beta}R^{\alpha\beta} - \frac13R^2\bigg) + \Lag_\text{matter}[g]\bigg] \,,
\end{align}
which will serve as the starting point for our analyses.

While it is possible to derive a fourth order action for the graviton directly from this action and quantize the resulting theory following the Ostrogradsky procedure for fourth order theories, it is perhaps more straightforward to expose the separate independent spin-2 DOFs already at the level of the Lagrangian. This can be achieved by expressing \eqref{SQG} as a second order theory through the introduction of an auxiliary field, $\Hab$, that will correspond to the extra hidden DOFs present in the fourth order version of the theory. Following the procedure laid out in \cite{Kubo:2022lja,Kubo:2022jwu}, this second order action is found to be given by
\begin{align} \label{Saux}
\Act_\text{aux} = \int\dx\gdet\bigg[\frac{\Mp^2}{2}R + \Mp G_{\alpha\beta}\iHab + \frac{\Mp^2}{4\alpha^2}\Big(\Hab\iHab - \Hs{\alpha}\Hs{\beta}\Big) + \Lag_\text{matter}[g]\bigg] \,,
\end{align}
which leads to the equation of motion (EOM)
\begin{align} \label{HabEOM}
\Hab = -\frac{\alpha^2}{\Mp}\left(2R_{\alpha\beta} - \frac{1}{3}\met R\right) \,,
\end{align}
that returns the fourth order action \eqref{SQG} after it is used to integrate $\Hab$ out of \eqref{Saux}.

We will be interested in co-moving tensor perturbations of the FLRW metric parameterized as\footnote{
  We suppress the scalar and vector parts of the metric perturbations as we are only interested in tensor perturbations here. We do note however, that there exists a contradictory debate on the question of whether, in the superhorizon regime, an undesirable growth of the scalar part of the ghost perturbation is a gauge artifact \cite{Deruelle:2010kf,Ivanov:2016hcm,Salvio:2017xul} or not \cite{DeFelice:2023psw}.} 
\begin{align} \label{FLRWpert}
\dd s^2 = a(\tau)^2\bar{g}_{\alpha\beta} \dd x^\alpha \dd x^\beta = a(\tau)^2\Big(-\dd\tau^2 + \big(\delta_{ij} + 2\Mp^{-1}h_{ij}\big)\dd x^i\dd x^j\Big) \,,
\end{align}
where $a(\tau)$ is the usual cosmological scale factor as a function of the conformal time $\tau$, which is related to the standard time coordinate by $\dd\tau=\dd t/a(t)$. It is thus convenient to re-express the action (\ref{Saux}) in terms of metric $\bar{g}_{\alpha\beta}$ defined above and a shifted auxiliary field $\bar{H}_{\alpha\beta}$:
\begin{align}
\Act_\text{aux} = \int\dd\tau\dd^3\bfx\sqrt{-\bar{g}}\bigg[&\frac{\Mp^2}{2}\Big(a^2R(\bar{g}) - 6a'^2\Big) + \Mp\bar{H}_{\alpha\beta}G^{\alpha\beta}(\bar{g}) \nnr
&+ \frac{\Mp^2}{4\alpha^2}\Big(\bar{H}_{\alpha\beta}\bar{H}^{\alpha\beta} - \bar{H}_{\alpha}{}^{\alpha}\bar{H}_{\beta}{}^{\beta}\Big) + a^4\Lag_\text{matter}[\bar{g}]\bigg] \,, \label{Saux2}
\end{align}
where we denote derivatives with respect to conformal time with a prime. This new tensor field is defined by
\begin{align} \label{bHDef}
\bar{H}_{\alpha\beta} = \Hab - \frac{\alpha^2}{\Mp}\bigg(\frac{4\bar{\nabla}_\beta\partial_\alpha a}{a} - \frac{8\partial_\alpha a\,\partial_\beta a}{a^2} + \frac{2\bar{g}_{\alpha\beta}\partial_\gamma a\,\partial^\gamma a}{a^2}\bigg) \,,
\end{align}
which is simply the original auxiliary field $\Hab$ minus its Weyl transformation, as dictated by \eqref{HabEOM}. Since the the sum of the second and third terms in the action (\ref{Saux}) is Weyl-scaling invariant, this quadratic part of our Weyl-rescaled auxiliary action has the exact same form as its counterpart in (\ref{Saux}). This leads to the crucial fact that, unlike the original auxiliary field $\Hab$, $\bar{H}_{\alpha\beta}$ functions as a perturbation with respect to the rescaled metric $\bar{g}_{\alpha\beta}$. One can see this explicitly by considering the EOM obtained by varying \eqref{Saux2} with respect to $\bar{H}_{\alpha\beta}$; this EOM has precisely the same form as \eqref{HabEOM}, except that the curvatures are evaluated using $\bar{g}_{\alpha\beta}$ as defined in \eqref{FLRWpert}. The right hand side of this EOM vanishes since $\bar{g}_{\alpha\beta}$ is just the Minkowski metric when $h_{ij}$ is set to zero, implying that $\bar{H}_{\alpha\beta}$ is not sourced at the background level and $\bar{H}_{\alpha\beta}=0$ is a solution of the classical theory at zeroth order in metric perturbations. Moving forward we will drop the bar and write $\Hab$ to avoid clutter, but it is important to remember that we are referring to the perturbation that appears in \eqref{Saux2} and not to the original auxiliary field introduced in \eqref{Saux} (which is sourced at the background level).


Moving forward, we may now expand \eqref{Saux2} to derive an action that governs the tensor perturbations. To this end, we note that both $h_{ij}$ and $H_{ij}$ satisfy the transverse-traceless conditions
\begin{align} \label{transverse}
&\partial_i h_{ij} = h_{ii} = 0  \,,& &\partial_i H_{ij} = H_{ii} = 0 \,,
\end{align}
which result, respectively, from fixing the gauge freedom associated with $h_{ij}$ and directly from the $H_{ij}$ EOM. Then, using that $\sqrt{-\bar{g}}R(\bar{g})=\big(-h_{ij}h''_{ij}+h_{ij}\nabla^2h_{ij}\big)+\dots$ and $G_{ij}=\big(h''_{ij}-\nabla^2h_{ij}\big)+\dots$, where $\dots$ stands for higher order terms in $h_{ij}$, we find that the quadratic part of \eqref{Saux2} describing tensor perturbations can be written as
\begin{align}
\Act_\text{tens}^{(0)}= \int\dd\tau\dd^3\bfx\bigg[&\frac{a^2}{2}\Big(h'_{ij}h'_{ij} + h_{ij}\nabla^2h_{ij}\Big) - \Big(h'_{ij}H'_{ij} + h_{ij}\nabla^2H_{ij}\Big) + \frac{\Mp^2}{4\alpha^2}H_{ij}H_{ij} \nnr
&+ h_{ij}h_{ij}\Big(a'^2 - 2a''a\Big) + \sqrt{-\bar{g}}\,a^4\Lag_\text{matter}[\bar{g}]\bigg] \,. \label{Stens0}
\end{align}
The matter part, $\sqrt{-\bar{g}}\,a^4\Lag_\text{matter}[\bar{g}]$, gives a contribution of $-(1/2)2a^2\Mp^{-1}T^t_{ij}$ to the Einstein EOM (i.e., the variation with respect to $h_{ij}$), where $T^t_{ij}$ is the transverse-traceless part of the energy momentum tensor given by
\begin{align}
T_{ij} = \bar{p}g_{ij} = -\Mp^2\bigg(\frac{2a''}{a} - \frac{a'^2}{a^2}\bigg)\big(\delta_{ij} + 2\Mp^{-1}h_{ij}\big) \,, \label{EMT}
\end{align}
and $\bar{p}$ is the unperturbed pressure of the background universe. Therefore, the relevant part of $\sqrt{-\bar{g}}\,a^4\Lag_\text{matter}$ is given by
\begin{align}
-\frac{a^2}{2\Mp}h_{ij}T^t_{ij} = h_{ij}\big(2a''a - a'^2\big)h_{ij} \,,
\end{align}
which leads to a handy cancellation of terms in the last line of (\ref{Stens0}).

Finally, to bring our action into the form that we will use as a basis moving forward, we rescale $H_{ij}$ as $H_{ij}\to a^2H_{ij}$ and shift $h_{ij}$ according to
\begin{align} \label{shift}
h_{ij} \to h_{ij}+H_{ij} \,,
\end{align}
to arrive at $\Act_\text{tens}=\int\dd\tau\dd^3\bfx\Lag_\text{tens}$, where
\begin{align}
\Lag_\text{tens} = a^2\bigg[&\frac12\big(h_{ij}'h_{ij}' - \nabla_k h_{ij}\nabla_k h_{ij}\big) - \frac12\big(H_{ij}'H_{ij}' - \nabla_k H_{ij}\nabla_k H_{ij}\big) \nnr
&- 2\frac{a'}{a}h_{ij}'H_{ij} + \bigg(\frac{m_\text{gh}^2a^2}{2} + \frac{a'^2}{a^2} + \frac{a''}{a}\bigg)H_{ij}H_{ij}\bigg] \label{Ltens0}
\end{align}
and we have identified the mass of the spin-2 ghost as
\begin{align} \label{ghost-mass}
m_\text{gh}=\frac{\Mp}{\sqrt{2}\alpha} \,.
\end{align}

\section{Quantum Theory in de Sitter space} \label{sec:canquant}

\subsection{Canonical quantization}

To proceed with our analyses we must next quantize our theory, which requires that we express our transverse perturbations in 3d-Fourier space according to
\begin{align} \label{Fdecomp}
\varphi_{ij}(\tau,\bfx) = \int\frac{\dd^3k}{(2\pi)^{3/2}}\sum_{\lambda=\pm2}\varphi_\lambda(\tau,\bfk)e_{ij}(\bfk/k,\lambda)e^{i \bfk\cdot\bfx} \,.
\end{align}
Here, $\varphi_{ij}$ represents an arbitrary tensor perturbation, $k=|\bfk|$ is the magnitude of the 3D momentum, $\lambda=\pm2$ is the helicity, and $e_{ij}(\bfk/k,\lambda)$ is a polarization tensor, which for $\bfk=(0,0,k)$, assumes the form
\begin{align}
e_{ij}(\bfk/k,\pm 2) = \frac{1}{2}\left(
  \begin{array}{ccc}1 & \pm i & 0\\
  \pm i & -1 & 0\\
  0 & 0 &0
  \end{array}\right) \,.
\end{align}

Using the decomposition \eqref{Fdecomp} and the identity $e_{ij}(\bfk/k,\lambda)e_{ij}(-\bfk/k,\lambda')=\delta_{\lambda \lambda'}$, we may establish the Fourier-transformed version of the action (\ref{Ltens0}) as $\hat{\Act}_\text{tens} = \int\dd\tau\dd^3\bfk\,\hat{\Lag}_\text{tens}$, where
\begin{align} \label{LtensF}
\hat{\Lag}_\text{tens} = \sum_{\lambda=\pm2}a^2(\tau)\bigg[&\frac{1}{2}\Big(h'_{\lambda}(\tau,\bfk)h'_{\lambda}(\tau,-\bfk) - k^2h_{\lambda}(\tau,\bfk)h_{\lambda}(\tau,-\bfk)\Big) \nnr
&-\frac{1}{2}\Big(H'_{\lambda}(\tau,\bfk)H'_{\lambda}(\tau,-\bfk) - k^2H_{\lambda}(\tau,\bfk)H_{\lambda}(\tau,-\bfk)\Big) - 2\frac{a'(\tau)}{a(\tau)}h_{\lambda}'(\tau,\bfk)H_{\lambda}(\tau,-\bfk) \nnr
&+ \bigg(\frac{m_\text{gh}^2a^2(\tau)}{2} + \frac{a'^2(\tau)}{a^2(\tau)} + \frac{a''(\tau)}{a(\tau)}\bigg)H_{\lambda}(\tau,\bfk)H_{\lambda}(\tau,-\bfk)\bigg]\,.
\end{align}
With this, we may define canonical momenta in the usual fashion,
\begin{align}
&p_{\lambda}(\tau,\bfk) = \frac{\partial{\hat{\Lag}}_\text{tens}}{\partial h'_{\lambda}(\tau,-\bfk)} = a^2(\tau)h'_{\lambda}(\tau,\bfk) - 2a(\tau)a'(\tau)H_{\lambda}(\tau,\bfk) \,, \label{mom-p} \\
&P_{\lambda}(\tau,\bfk) = \frac{\partial\hat{\Lag}_\text{tens}}{\partial H'_{\lambda}(\tau,-\bfk)} = -a^2(\tau)H'_{\lambda}(\tau,\bfk)
\label{mom-P} \,,
\end{align}
and subsequently obtain the Hamiltonian 
\begin{align} \label{hamiltonian}
\hat{\Ham} &= \int\dd^3\bfk\sum_{\lambda = \pm 2}\Big[p_{\lambda}(\tau,\bfk)h'_{\lambda}(\tau,-\bfk) + P_{\lambda}(\tau,\bfk)H'_{\lambda}(\tau,-\bfk) - \hat{\Lag}_\text{tens}\Big] \nnr
&= \int\dd^3\bfk\sum_{\lambda=\pm2}\bigg[\frac{1}{2}\Big(a^{-2}(\tau)p_{\lambda}(\tau,\bfk)p_{\lambda}(\tau,-\bfk) + a^2k^2h_{\lambda}(\tau,\bfk)h_{\lambda}(\tau,-\bfk)\Big)\nnr
&\phantom{= \int\dd^3\bfk\,}- \frac{1}{2}\Big(a^{-2}(\tau)P_{\lambda}(\tau,\bfk)P_{\lambda}(\tau,-\bfk) + a^2k^2H_{\lambda}(\tau,\bfk)H_{\lambda}(\tau,-\bfk)\Big) + 4\frac{a'(\tau)}{a(\tau)}p_{\lambda}(\tau,\bfk)H_{\lambda}(\tau,-\bfk) \nnr
&\phantom{= \int\dd^3\bfk\,} - a^2(\tau)\bigg(\frac{m_\text{gh}^2a^2(\tau)}{2} - \frac{a'^2(\tau)}{a^2(\tau)} + \frac{a''(\tau)}{a(\tau)}\bigg)H_{\lambda}(\tau,\bfk)H_{\lambda}(\tau,-\bfk)\bigg] \,.
\end{align}
Finally, the quantum theory is established after postulating the canonical equal-time commutation relations
\begin{align} \label{CCR}
\big[h_\lambda(\tau,\bfk),p_{\lambda'}(\tau,\bfk')\big] = \big[H_\lambda(\tau,\bfk),P_{\lambda'}(\tau,\bfk')\big] = i\delta_{\lambda\lambda'}\delta^3(\bfk + \bfk') \,,
\end{align}
noting that all other commutators vanish.

\subsection{The diagonalized canonical system} \label{sec:diag-con}

The equations of motion (EOMs) that follow from our second-order Lagrangian \eqref{Ltens0} are given by
\begin{align}
&h''_{ij} - \nabla^2h_{ij} + 2\frac{a'}{a}\big(h'_{ij} - H'_{ij}\big) - 2\bigg(\frac{a'^2}{a^2} + \frac{a''}{a}\bigg)H_{ij} = 0 \,, \label{eom-h} \\
&H''_{ij} - \nabla^2H_{ij} - 2\frac{a'}{a}\big(h'_{ij} - H'_{ij}\big) + 2\bigg(\frac{a'^2}{a^2} + \frac{a''}{a} + \frac{m_\text{gh}^2a^2}{2}\bigg)H_{ij} = 0 \,, \label{eom-H}
\end{align}
and one may combine them to further obtain the equations
\begin{align}
&\big(h_{ij} + H_{ij}\big)'' - \nabla^2\big(h_{ij} + H_{ij}\big) + m^2a^2H_{ij} = 0 \,, \label{eom-hH} \\
&\big(h_{ij} + H_{ij}\big)'''' - 2\nabla^2\big(h_{ij} + H_{ij}\big)'' + \nabla^4\big(h_{ij} + H_{ij}\big) \nnr
&\qquad+ m_\text{gh}^2a^2\bigg(\big(h_{ij} + H_{ij}\big)'' - \nabla^2\big(h_{ij} + H_{ij}\big) + 2\frac{a'}{a}\big(h_{ij} + H_{ij}\big)'\bigg) = 0 \,. \label{4th-eom}
\end{align}
The last equation above matches the EOM that is derived from the fourth-order expression of linearized QG that is more often employed, where one does not separate the massless and massive spin-2 modes at the level of the action as we have. 

It is known that the fourth-order de Sitter spacetime EOM (\ref{4th-eom}) admits two independent solutions and that each of them satisfies its own second-order EOM \cite{Clunan:2009er,Salvio:2017xul}, which are obtained by factoring the fourth-order equation. We note that these are distinct from the EOMs \eqref{eom-h} and \eqref{eom-H} that we obtain in our formulation. In momentum space, these new EOMs can be written as
\begin{align}
&\ssz{h}''_{\lambda}(\tau,\bfk) + k^2\ssz{h}_{\lambda}(\tau,\bfk) - \frac{2}{\tau}\ssz{h}'_{\lambda}(\tau,\bfk) = 0 \,, \label{eom-h0} \\
&\ssz{H}''_{\lambda}(\tau,\bfk) + k^2\ssz{H}_{\lambda}(\tau,\bfk) - \frac{2}{\tau}\ssz{H}'_{\lambda}(\tau,\bfk) + \frac{1}{\rho\tau^2}\bigg(1 + 2\rho\bigg)\ssz{H}_{\lambda}(\tau,\bfk) = 0 \,, \label{eom-H0}
\end{align}
where $\ssz{h}_{\lambda}$ and $\ssz{H}_{\lambda}$ are the solutions to this alternate set of diagonal EOMs, $\rho=\bm{H}^2/m_\text{gh}^2$, $\bm{H}$ is the Hubble rate, and we have used the pure de Sitter spacetime expression for the scale factor:
\begin{align} \label{scalefactor}
a(\tau) = -\frac{1}{\bm{H}\tau} \,.
\end{align}

Given the fact that ($h_{\lambda} + H_{\lambda}$) and ($\ssz{h}_{\lambda} + \ssz{H}_{\lambda}$) satisfy the same fourth-order EOM (\ref{4th-eom}), we assume that the identity
\begin{align}
h_{\lambda} + H_{\lambda} = \ssz{h}_{\lambda} + \ssz{H}_{\lambda}
\label{h+H}
\end{align}
holds, and that the canonical momenta for the diagonal fields are given by
\begin{align}
&\ssz{p}_{\lambda}(\tau,\bfk) = a^2(\tau)\ssz{h}'_{\lambda}(\tau,\bfk) \,,& &\ssz{P}_{\lambda}(\tau,\bfk) = -a^2(\tau)\ssz{H}'_{\lambda}(\tau,\bfk) \,. \label{mom-pP0}
\end{align}
It should also be noted that since the EOMs \eqref{eom-h0} and \eqref{eom-H0} for $\tilde{h}$ and $\tilde{H}$ are linear, their normalization is not fixed at this point. Therefore, \eqref{h+H} is the most general expression under the assumption that $h+H$ can be written as a linear combination of $\tilde{h}$ and $\tilde{H}$. Then, from (\ref{eom-hH}), (\ref{h+H}), and the fact that we can write $h_{\lambda}=(h_{\lambda}+H_{\lambda})-H_{\lambda}=(\ssz{h}_{\lambda}+\ssz{H}_{\lambda})-H_{\lambda}$, we can express the original canonical coordinates as
\begin{align}
&h_{\lambda} = \ssz{h}_{\lambda} - 2\rho\,\ssz{H}_{\lambda} + 2\bm{H}^2\rho\tau^3\big(\ssz{p}_{\lambda} - \ssz{P}_{\lambda}\big) \,, \label{h-lambda} \\
&H_{\lambda} = (1 + 2\rho)\ssz{H}_{\lambda} - 2\bm{H}^2\rho\tau^3\big(\ssz{p}_{\lambda} - \ssz{P}_{\lambda}\big) \label{H-lambda} \,.
\end{align}
Similar calculations also yield analogous expressions of the canonical momenta:
\begin{align}
&p_{\lambda} = (1 + 2\rho)\ssz{p}_{\lambda} - \frac{2k^2\rho}{\bm{H}^2\tau}\big(\ssz{h}_{\lambda} + \ssz{H}_{\lambda}\big) \,, \label{p-lambda} \\
&P_{\lambda}(\tau, \bfk) = 6\rho\,\ssz{p}_{\lambda} + (1 - 4\rho)\ssz{P}_{\lambda} - \frac{2k^2\rho}{\bm{H}^2\tau}\ssz{h}_{\lambda} - \frac{2\big(1 + \rho(2+k^2\tau^2)\big)}{\bm{H}^2\tau^3}\ssz{H}_{\lambda} \,. \label{P-lambda}
\end{align}
Putting all of the above together, one can use the relations (\ref{mom-pP0}--\ref{P-lambda}) with the EOMs (\ref{eom-h0}) and (\ref{eom-H0}) to explicitly 
show that $h_{\lambda}$ and $H_{\lambda}$ as given in (\ref{h-lambda}) and (\ref{H-lambda}) satisfy, respectively, (\ref{eom-h}) and (\ref{eom-H}) when rewritten 
in terms of Fourier modes on de Sitter spacetime.

It is also possible to express the diagonal fields in terms of their original off-diagonal counterparts. This is most easily done by bringing the relations (\ref{h-lambda}--\ref{P-lambda}) into matrix form:
\begin{align}
&\left(
  \begin{array}{c}\
  h_{\lambda} \\
  p_{\lambda} \\
  H_{\lambda} \\
  P_{\lambda}
  \end{array} \right) = {\cal C} \left(
  \begin{array}{c}
  \ssz{h}_{\lambda} \\
  \ssz{p}_{\lambda} \\
  \ssz{H}_{\lambda} \\
  \ssz{P}_{\lambda}
  \end{array}\right) \,,
\end{align}
where
\begin{align} \label{matrix-C}
{\cal C} = \left(
  \begin{array}{cccc}
  1 & 2 \bm{H}^2\rho  \tau^3 & -2 \rho & -2 \bm{H}^2 \rho \tau^3 \\
  -2 k^2 \rho/\bm{H}^2 \tau & 1 + 2 \rho & -2 k^2 \rho/\bm{H}^2\tau & 0 \\
  0 & -2 \bm{H}^2 \rho \tau^3 & 1+2\rho & 2 \bm{H}^2 \rho \tau^3 \\
  -2 k^2 \rho/\bm{H}^2 \tau & 6 \rho & -2 \big(1+\rho(2 +k^2 \tau^2)\big)/\bm{H}^2 \tau^3 & 1-4\rho
  \end{array}\right) \,.
\end{align}
Then, using the inverse of ${\cal C}$, 
\begin{align}
{\cal C}^{-1} = \frac{1}{(1+ 2 \rho)}\left(
  \begin{array}{cccc}
  1+2 \rho  & -2  \bm{H}^2\rho  \tau^3 & 6 \rho & 2 \bm{H}^2 \rho \tau^3 \\
  2 k^2 \rho/\bm{H}^2 \tau & 1  &   2 k^2 \rho/\bm{H}^2\tau & 0 \\
  0 & 2 \bm{H}^2 \rho \tau^3 & 1-4 \rho &  -2 \bm{H}^2 \rho \tau^3 \\
  2 k^2 \rho/\bm{H}^2 \tau & -2 \rho & 2 \big(1+\rho(2 +k^2 \tau^2)\big)/\bm{H}^2 \tau^3 & 1+2\rho
  \end{array}\right) \,,
\end{align}
we find the following non-zero equal-time commutation relations between $\ssz{h}_{\lambda}$, $\ssz{p}_{\lambda}$, $\ssz{H}_{\lambda}$, and $\ssz{P}_{\lambda}$:
\begin{align}
&\big[\ssz{h}_{\lambda}(\tau,\bfk),\ssz{p}_{\lambda'}(\tau,\bfk')\big] = \frac{i\delta_{\lambda\lambda'}\delta^3(\bfk + \bfk')}{1 + 2\rho} \,,& &\big[\ssz{H}_{\lambda}(\tau,\bfk),\ssz{P}_{\lambda'}(\tau,\bfk')\big] = \frac{i\delta_{\lambda\lambda'}\delta^3(\bfk + \bfk')}{1 + 2\rho} \,. \label{CCR-hH0}
\end{align}
These relations imply that (\ref{matrix-C}) is a canonical transformation, at least up to the normalization factor $(1+2\rho)^{-1}$. Using the ``inverse'' of the expressions (\ref{CCR-hH0}), we can write
\begin{align}
&\!\!\!\big[h_\lambda(\tau,\bm k),p_{\lambda'}(\tau,\bm k')\big] = \big(1 + 2\rho + 4k^2\rho^2\tau^2\big)\big[\ssz{h}_\lambda(\tau,\bm k),\ssz{p}_{\lambda'}(\tau,\bm k')\big] - 4k^2\rho^2\tau^2\big[\ssz{H}_\lambda(\tau,\bm k),\ssz{P}_{\lambda'}(\tau,\bm k')\big] \label{CCR-h} \\
&\!\!\!\big[H_\lambda(\tau,\bm k),P_{\lambda'}(\tau,\bm k')\big] = -4k^2\rho^2\tau^2\big[\ssz{h}_\lambda(\tau,\bm k),\ssz{p}_{\lambda'}(\tau,\bm k')\big] + \big(1 + 2\rho + 4k^2\rho^2\tau^2\big)\big[\ssz{H}_\lambda(\tau,\bm k),\ssz{P}_{\lambda'}(\tau,\bm k')\big] \label{CCR-H}
\end{align}
which allows us to derive the summed commutator
\begin{align}
&\big[h_{\lambda}(\tau,\bfk) + H_{\lambda}(\tau,\bfk),p_{\lambda'}(\tau,\bfk') + P_{\lambda'}(\tau, \bfk')\big] \nnr
&\qquad = (1 + 2\rho)\Big(\big[\ssz{h}_{\lambda}(\tau,\bfk),\ssz{p}_{\lambda'}(\tau,\bfk')\big] + \big[\ssz{H}_{\lambda}(\tau,\bfk),\ssz{P}_{\lambda'}(\tau,\bfk')\big]\Big) \nnr
&\qquad = (1 + 1)i\delta_{\lambda\lambda'}\delta^3(\bfk + \bfk') \,. \label{CCR-h+H}
\end{align}
Here, ``$(1+1)$'' comes from the sum of contributions from the canonical commutation relations between $h_{\lambda}(\tau,\bfk)$ and $p_{\lambda'}(\tau,\bfk')$, and
between $H_{\lambda}(\tau,\bfk)$ and $P_{\lambda'}(\tau,\bfk')$, which can be understood as a sum of the contributions from both the massless and massive modes. In the upcoming sections when discussing quasi de Sitter spacetime, we will consider only the massless mode and assume that this interpretation remains valid.

\subsection{Exact solutions and the power spectrum} \label{sec:exactsols}

As previously mentioned, the exact solutions of the diagonal EOMs (\ref{eom-h0}) and (\ref{eom-H0}) are known \cite{Starobinsky:1979ty,Vilenkin:1982wt,Linde:1982uu,Rubakov:1982df,Kodama:1984ziu}, and they may be expressed as functions of $z=-k\tau$: 
\begin{align}
&\ssz{h}_{\lambda}(\tau,\bfk) = \!-\bm{H}\tau\Big(s_h(z)\ssz{\alpha}_{\lambda}(\bfk) + s_h^*(z)\ssz{\alpha}_{\lambda}^\dag(-\bfk)\!\Big) \,,& &\!\!\!\ssz{H}_{\lambda}(\tau,\bfk) = \!-\bm{H}\tau\Big(s_H(z)\ssz{\beta}_{\lambda}(\bfk) + s_H^*(z)\ssz{\beta}_{\lambda}^\dag(-\bfk)\!\Big) \,, \label{exact-hH0}
\end{align}
where
\begin{align}
&s_h(z) = (1/z - i)e^{i z} \,,& &s_H(z) = \Big(\frac{\pi z}{2}\Big)^{1/2}e^{i\pi(\nu/2 + 1/4)}\Big(J_\nu(z) + iY_\nu(z)\Big) \label{shH} \,.
\end{align}
Here, $J_\nu$ and $Y_\nu$ are the Bessel function of the first and second kind with order $\nu=(1/4-1/\rho)^{1/2}$, and the massless and massive mode functions are given by $\ssz{h}_{\lambda}$ and $\ssz{H}_{\lambda}$, respectively. We have also factored out a factor of $-\bm{H}\tau$ $(\approx 1/a(\tau))$ in (\ref{exact-hH0}) for future convenience, as we will consider $h_{\lambda}=\bar{h}_{\lambda}/a(\tau)$ and $H_{\lambda}=\bar{H}_{\lambda}a(\tau)$ when discussing the case of quasi de Sitter spacetime in the next section, where we fix the normalization of the asymptotic solution for $\bar{h}_{\lambda}$ and $\bar{H}_{\lambda}$ using $s_h$ and $s_H$. We also note that the order parameter $\nu$ is real (imaginary) for $\rho=\bm{H}^2/m^2 > (<) 4$.

The canonical momenta can be calculated from (\ref{mom-pP0}), leading to the commutation relations
\begin{align}
&\big[\ssz{\alpha}_{\lambda}(\bfk),\ssz{\alpha}^\dag_{\lambda'}(-\bfk')\big] = -\frac{i}{2k}\big[\ssz{h}_{\lambda}(\tau,\bfk),\ssz{p}_{\lambda'}(\tau,\bfk')\big] = \frac{\delta_{\lambda\lambda'}\delta^3(\bfk + \bfk')}{2k(1+2 \rho)} \,, \label{CCR-alpha0} \\
&\big[\ssz{\beta}_{\lambda}(\bfk),\ssz{\beta}^\dag_{\lambda'}(-\bfk')\big] = \frac{i}{2k}\big[\ssz{H}_{\lambda}(\tau,\bfk),\ssz{P}_{\lambda'}(\tau,\bfk')\big] = -\frac{\delta_{\lambda\lambda'}\delta^3(\bfk + \bfk')}{2k(1+2 \rho)} \,, \label{CCR-beta0}
\end{align}
where we note that the last relation (\ref{CCR-beta0}) is valid for both real and imaginary $\nu$\footnote{
  If the factor $e^{i\pi\nu/2}$ in the second part of (\ref{shH}) is omitted and $\nu=i\bar{\nu}$ is imaginary, the last term of (\ref{CCR-beta0}) should be multiplied with $e^{-\pi\bar{\nu}}$, however, this change of the normalization does not alter the power spectrum \eqref{PS-H0-2}.}.
Using the relations above, we can finally calculate the power spectrum in pure de Sitter spacetime. To do so we first define the vacuum according to \cite{Bunch:1978yq}:
\begin{align}
&\ssz{\alpha}_{\lambda}(\bfk)\big|0\big\rangle = 0 \,,& &\ssz{\beta}_{\lambda}(\bfk)\big|0\big\rangle = 0 \,,
\label{vacuum}
\end{align}
and then, using $\ssz{h}_{\lambda}$ as given in (\ref{exact-hH0}) with (\ref{shH}) and the commutation relation (\ref{CCR-alpha0}), we find that for the massless mode we have
\begin{align}
{\cal P}_{\ssz{h}\ssz{h}}({\bm k,\bm k'}) &= \frac{4}{\Mp^2}\sum_{\lambda=\pm}\mel{0}{\ssz{h}_{\lambda}(\tau,\bfk)\ssz{h}_{\lambda}(\tau,\bfk')}{0} = \frac{4\bm{H}^2\big(1 + k^2\tau^2\big)}{\Mp^2k^3(1 + 2\rho)}\delta^3(\bfk + \bfk') \label{PS-h0-1} \\
&\approx \frac{4\bm{H}^2}{\Mp^2k^3\,(1 + 2\rho)}\delta^3(\bfk + \bfk') \,, \label{PS-h0-2}
\end{align}
where the last equality is understood to be in the asymptotic ($-\tau\to0$) limit.

The massive mode should be treated separately, depending on whether $\nu=\sqrt{1/4-1/\rho}$ is real $(\rho>4)$ or imaginary $(\rho<4)$. First we consider the massive mode with a real $\nu$ in the asymptotic limit where $-k\tau\ll1$:
\begin{align}
{\cal P}_{\ssz{H}\ssz{H}}({\bm k,\bm k'}) = \frac{4}{\Mp^2}\sum_{\lambda=\pm}\mel{0}{\ssz{H}_{\lambda}(\tau,\bfk)\ssz{H}_{\lambda}(\tau,\bfk')}{0} \approx -\frac{2^{2\nu+1}\bm{H}^2(-k\tau)^{3 - 2\nu}\,\Gamma^2(\nu)}{\pi\Mp^2k^3(1 + 2\rho)}\delta^3(\bfk + \bfk') \label{PS-H0-2} \,,
\end{align}
where we have used the identity $\Gamma(1-\nu)\Gamma(\nu)=\pi/\sin(\nu\pi)$ to obtain the asymptotic formula for $z\ll1$:
\begin{align}
J_\nu(z) + iY_\nu(z) \approx iY_\nu(z) &= -i\frac{2^\nu z^{-\nu}}{\sin(\nu\pi)\Gamma(1 - \nu)}(1 + O(z^2)) \nnr
&\approx -i\frac{2^\nu\Gamma(\nu)}{\pi}z^{-\nu} \,. \label{J+Y-1}
\end{align}
In the case of an imaginary $\nu=i\bar{\nu}=i\sqrt{1/\rho-1/4}$, the asymptotic formula (\ref{J+Y-1}) changes to
\begin{align}
J_{i\bar{\nu}}(z) + iY_{i\bar{\nu}}(z) &\approx -i\frac{2^{-i\bar{\nu}}}{\bar{\nu}\Gamma(i\bar{\nu})}z^{i\bar{\nu}} + i\bigg(\frac{2^{i\bar{\nu}}\csch(\bar{\nu}\pi)}{\bar{\nu}\Gamma(-i\bar{\nu})}z^{-i\bar{\nu}} - \frac{2^{-i\bar{\nu}}\coth(\bar{\nu}\pi)}{\bar{\nu}\Gamma(i\bar{\nu})}z^{i\bar{\nu}}\bigg) \nnr
&\approx -i\frac{2^{-i\bar{\nu} + 1}}{\bar{\nu}\Gamma(i\bar{\nu})}z^{i\bar{\nu}} \,, \label{J+Y-2}
\end{align}
where in the last line we assumed both $|z|\ll1$ and large $\bar{\nu}$. With these considerations, we obtain the massive mode power spectrum for imaginary $\nu$:
\begin{align}
{\cal P}_{\ssz{H}\ssz{H}}({\bm k,\bm k'}) \approx -\frac{8\pi\bm{H}^2(-k\tau)^{3}e^{-\bar{\nu}\pi}}{\Mp^2\bar{\nu}^2k^3|\Gamma(i\bar{\nu})|^2(1 + 2\rho)}\delta^3(\bfk + \bfk') \approx -\frac{4\bm{H}^2\rho^{1/2}(-k\tau)^{3}}{\Mp^2k^3(1+2\rho)}\delta^3(\bfk+\bfk') \,,
\label{PS-H0-3}
\end{align}
where we have used that $|\Gamma(i\bar{\nu})|^2\approx2\pi e^{-\bar{\nu}\pi}/\bar{\nu}$ for large $\bar{\nu}=(1/\rho-1/4)^{1/2}\approx\rho^{-1/2}$. It is important to note that the power spectrum is suppressed by $\rho^{1/2}=\bm{H}/m$, which should be compared with the exponential suppression derived in \cite{Pilo:2004ke,Bartolo:2004if}.

As we see from (\ref{PS-H0-2}) and (\ref{PS-H0-3}), the power spectrum for the massive mode depends on $\tau$ and decreases as $(-\tau)$ approaches zero, while the massless mode spectrum (\ref{PS-h0-2}) approaches a constant in this same limit. Furthermore, as we can see from (\ref{J+Y-1}) and (\ref{J+Y-2}), the solution proportional to $Y_\nu(z)$ dominates in the asymptotic limit if $\nu$ is real, while for an imaginary $\nu$, the solutions proportional to $Y_\nu(z)$ and $J_\nu(z)$ are equally relevant.

Finally, to establish the validity of our second-order formulation, we may derive the power spectrum corresponding to the combined perturbation $\ssz{h}+\ssz{H}=h+H$. In light of (\ref{h+H}), this can be expressed as the sum of (\ref{PS-h0-2}) and (\ref{PS-H0-3}) after neglecting the massive mode, which dies in the asymptotic limit for both the real and imaginary ($\rho> 4$ and $\rho<4$) cases. With these considerations, we find
\begin{align}
{\cal P}_{hH}({\bm k,\bm k'}) &= \frac{4}{\Mp^2}\sum_{\lambda=\pm}\mel{0}{\big(h_{\lambda}(\tau,\bfk) + H_{\lambda}(\tau,\bfk)\big)\big(h_{\lambda}(\tau,\bfk') + H_{\lambda}(\tau,\bfk')\big)}{0} \nnr
&= \frac{4\bm{H}^2}{\Mp^2k^3(1+2\rho)}\delta^3(\bfk + \bfk') \label{PS-hH0-2}
\end{align}
in the asymptotic limit, which agrees with the known results of \cite{Clunan:2009er,Myung:2014jha,Myung:2015vya,Salvio:2017xul,Anselmi:2020lpp} and reduces to the result of General Relativity in the $\rho={\bm H}^2/m_\text{gh}^2\to0$ limit \cite{Starobinsky:1979ty}.

\section{Tensor perturbations on quasi de Sitter} \label{sec:quasi-dS}

\subsection{Slow-rolling equations of motion and asymptotic solutions}

Now that we have established how our formalism may be employed to calculate the power spectrum of tensor perturbations on exact de Sitter spacetime, we will now consider the more physically realistic case of quasi de Sitter spacetime where the Hubble rate $\bm{H}$ is not a constant, but varies slightly with $\tau$. This may be quantified by writing
\begin{align}
\frac{\dd\bm{H}(\tau)}{\dd t} = a({\tau})^{-1}\bm{H}'(\tau) = -\epsilon\,\bm{H}^2(\tau) \,,
\label{dotH}
\end{align}
where $\epsilon \lesssim 0.004$ is a slow-roll parameter \cite{Planck:2018jri,BICEP:2021}. In the following discussions we will treat $\epsilon$ as a small perturbation parameter and will keep only terms up to and including $O(\epsilon)$, neglecting higher order terms. Additionally, though $\epsilon$ actually depends on $\tau$, we ignore its $\tau$ dependence because it does not appear until $O(\epsilon^2)$. Further, from $(a\bm{H})'=(1-\epsilon)(a\bm{H})^2$ it follows that
\begin{align}
a(\tau)\bm{H}(\tau) = -\frac{1}{\tau(1 - \epsilon)} + O(\epsilon^2) \approx -\frac{1 + \epsilon}{\tau} \,. \label{aH}
\end{align}
Similarly, one also finds the relations
\begin{align}
&\rho'(\tau) = \frac{\dd\bm{H}^2(\tau)/\dd\tau}{m_\text{gh}^2} \approx 2\epsilon\,\rho(\tau)/\tau \,,& &\rho''(\tau) \approx -2\epsilon\,\rho(\tau)/\tau^2 \,. \label{rho-dash}
\end{align}

Moving forward, it will be more convenient to consider the re-scaled fields $\bar{h}_\lambda(\tau,\bfk)=a(\tau)h_\lambda(\tau,\bfk)$ and $\bar{H}_\lambda(\tau,\bfk)=a(\tau)H_\lambda(\tau,\bfk)$, as previously mentioned. Furthermore, since only the massless mode survives in the asymptotic limit, we can safely drop the $\beta_\lambda$ mode operators and assume that the rescaled fields take the forms
\begin{align}
\bar{h}_\lambda(\tau,\bfk) &= f(z)\alpha_\lambda(\bfk) + f^*(z)\alpha_\lambda^\dag(-\bfk) = f^R(z)\Big(\alpha_\lambda(\bfk) + \alpha_\lambda^\dag(-\bfk)\Big) + if^I(z)\Big(\alpha_\lambda(\bfk) - \alpha_\lambda^\dag(-\bfk)\Big) \,, \label{h-bar} \\
\bar{H}_\lambda(\tau,\bfk) &= g(z)\alpha_\lambda(\bfk) + g^*(z)\alpha_\lambda^\dag(-\bfk) = g^R(z)\Big(\alpha_\lambda(\bfk) + \alpha_\lambda^\dag(-\bfk)\Big) + ig^I(z)\Big(\alpha_\lambda(\bfk) - \alpha_\lambda^\dag(-\bfk)\Big) \,, \label{H-bar}
\end{align}
where $f^R(z)$, $f^I(z)$, $\dots$ are real functions of $z=-k\tau$. Using the EOMs (\ref{eom-h}) and (\ref{eom-H}), we find that these functions should satisfy
\begin{align}
&\bigg(\frac{\dd^2}{\dd z^2} - \frac{2 + 3\epsilon}{z^2} + 1\bigg)f^{R,I}(z) + \frac{2}{z}\bigg((1 + \epsilon)\frac{\dd}{\dd z} - \frac{2 + 3\epsilon}{z}\bigg)g^{R,I}(z) + O(\epsilon^2) = 0 \,, \label{eom-f} \\
&\bigg(\frac{\dd^2}{\dd z^2} + \frac{(1+2\epsilon)/\rho + 7\epsilon + 4}{z^2} + 1\bigg)g^{R,I}(z) + \frac{2}{z}\bigg((1 + \epsilon)\frac{\dd}{\dd z} + \frac{1 + 2\epsilon}{z}\bigg)f^{R,I}(z) + O(\epsilon^2) = 0 \,, \label{eom-g}
\end{align}
in quasi de Sitter spacetime, where we have used $a'/a \approx -(1+\epsilon)/\tau$ and $a''/a \approx (2+3\epsilon)/\tau^2$.

We now look for asymptotic solutions for the coupled system of DEs above and restrict ourselves to the massless mode i.e.\ we look for an $O(\epsilon)$ correction to $h_{\lambda}(\tau,\bfk)$ as defined in (\ref{h-lambda}). We will use the fact that knowledge of the asymptotic solution is sufficient for calculating an $O(\epsilon)$ correction to the commutation relation (\ref{CCR-alpha0}) since this relation is $\tau$-independent. Additionally, we will employ the solution at $\epsilon=0$ to fix the normalization of the asymptotic solution at $\epsilon=0$ and we will require that the asymptotic solutions of (\ref{eom-f}) and  (\ref{eom-g}) at $O(\epsilon)$ reduce to the corresponding solutions at $\epsilon=0$\footnote{
  This does not fix the normalization  at $O(\epsilon)$. We will come to discuss this point later on.}.

Using the canonical transformation (\ref{matrix-C}), we  find that $f(z)$ for the massless mode at $\epsilon=0$ is given by
\begin{align}
f(z) = s_h(z) + \frac{2\rho\tau}{(-\bm{H}\tau)}\frac{\dd}{\dd\tau}\Big((-\bm{H}\tau) s_h(z)\Big) = \frac{1}{z}\Big(1 - iz + 2\rho z^2\Big)e^{i z} \,, \label{f0}
\end{align}
where $s_h(z)$ is defined in (\ref{shH}). With this, the real and imaginary parts of $f(z)$ at $\epsilon=0$ are straightforwardly found to be
\begin{align}
f^R(z) &= \Re f(z) = \sin(z) + \big(z^{-1} + 2\rho z\big)\cos(z) \nnr
&= z^{-1} + (1/2 + 2\rho)z - (1/8 + \rho)z^3 + O(z^5) \,, \label{fR0} \\
f^I(z) &= \Im f(z) =  \big(z^{-1} + 2\rho z\big)\sin(z) - \cos(z) \nnr
&= (1/3 + 2\rho)z^2 - (1/30 + \rho/3)z^4 + O(z^6) \,. \label{fI0}
\end{align}
Similarly for $g(z)=g^R(z)+ig^I(z)$, we find
\begin{align}
g^R(z) &= \Re g(z) = -2\rho z\cos(z) \nnr
&= \rho \Big(-2z + z^3 - z^5/12 + O(z^7)\Big) \,, \label{gR0} \\
g^I(z) &= \Im g(z)  = -2\rho z\sin(z) \nnr
&= \rho\Big(-2z^2 + z^4/3 - z^6/60 + O(z^8)\Big) \,. \label{gI0}
\end{align}

These asymptotic solutions at $\epsilon=0$, (\ref{fR0}--\ref{gI0}), suggest the following ansatz for an $O(\epsilon)$ solution:
\begin{align}
 &f_R(z) = c_1\,z^{\gamma_1} + c_2\,z^{\gamma_1 + 2} + c_3\,z^{\gamma_1 + 4} + \dots& &f_I(z) = \hat{c}_1\,z^{\hat{\gamma}_1} + \hat{c}_2\,z^{\hat{\gamma}_1 + 2} + \hat{c}_3\,z^{\hat{\gamma}_1 + 4} + \dots \,, \label{ansatz-f} \\
 &g_R(z) = d_1\,z^{\gamma_2} + d_2\,z^{\gamma_2 + 2} + \dots& &g_I(z) = \hat{d}_1\,z^{\hat{\gamma}_2} + \hat{d}_2\,z^{\hat{\gamma}_2 + 2} + \hat{d}_3\,z^{\hat{\gamma}_2 + 4} + \dots \,. \label{ansatz-g}
\end{align}
Inserting this ansatz into the DEs (\ref{eom-f}) and (\ref{eom-g}), dropping terms $O(\epsilon^2)$, and solving for the coefficients yields
\begin{align}
&c_1 = 1& &\hat{c}_1 =\frac{1}{3} + 2\rho + \epsilon\,\frac{1 + 12\rho}{9(1 + 2\rho)} \nnr
&c_2 = \frac{1 + 4\rho}{2} - \epsilon\,(1 + 6\rho)& &\hat{c}_2 = -\frac{3 + 10\rho}{30} + \epsilon\,\frac{1 + 112\rho(1 + 15\rho)}{450(1 + 2\rho)(1 + 12\rho)} \nnr
&c_3 =-\frac{1 + 8\rho}{8} + \epsilon\,\frac{2\rho(1 + 9\rho)}{1 + 6\rho} \qquad\qquad& &\hat{c}_3 =\frac{1 + 14\rho}{840} + \epsilon\,\frac{37 + 3542\rho + 48\rho^2(2227 + 20370\rho)}{88200D} \label{c3def} \\
&d_1 = -2\rho + \epsilon\,6\rho& &\hat{d}_1 = -2\rho \nnr
&d_2 = \rho - \epsilon\,\frac{\rho(1 + 18\rho)}{1 + 6\rho}& &\hat{d}_2 =\frac{\rho}{3} -\epsilon\,\frac{2\rho(1 + 24\rho)}{9(1 + 14\rho + 24\rho^2)} \nnr
&\phantom{d_3 = 0} & &\hat{d}_3 = -\frac{\rho}{60} + \epsilon\,\frac{4\rho(1 + 57\rho + 720\rho^2)}{225D} \nnr
&\gamma_1 = -1 -\epsilon& &\hat{\gamma}_1 = \hat{\gamma}_2 = 2 -\epsilon\,\frac{1 - 2\rho}{1 + 2\rho} = 2 + \epsilon\,\delta\hat{\gamma} \label{gammadef} \\
&\gamma_2 = 1 - \epsilon \,, \nonumber 
\end{align}
where $D=(1+2\rho)(1+12\rho)(1+30\rho)$, thus establishing our desired $O(\epsilon)$ solution.

\subsection{The $O(\epsilon)$ power spectrum}

Finally, with the above solution in hand, we are position to calculate the power spectrum at $O(\epsilon)$. To do so, we must first fix the normalization of the commutation relation between the massless mode operators $\alpha_\lambda(\bfk)$ and $\alpha^\dag_\lambda(\bfk')$ by using the equal-time canonical commutation relations (\ref{CCR}).
This requires that we compute the corresponding canonical momenta in quasi de Sitter:
\begin{align}
p_\lambda(\tau,\bfk) &= a(\tau)\bar{p}_\lambda(\tau,\bfk) = a(\tau)^2\bigg(\frac{\bar{h}_\lambda(\tau,\bfk)}{a(\tau)}\bigg)' - 2a'(\tau)\bar{H}_\lambda(\tau,\bfk) \nnr
&=-ka(\tau)\bigg(\frac{\dd\bar{h}_\lambda(\tau,\bfk)}{\dd z} + \frac{1}{z}(1 + \epsilon)\big(\bar{h}_\lambda(\tau,\bfk) + 2\bar{H}_\lambda(\tau,\bfk)\big)\bigg) \,, \\
P_\lambda(\tau,\bfk) &= a(\tau)\bar{P}_\lambda(\tau,\bfk) = -a(\tau)^2\bigg(\frac{\bar{h}_\lambda(\tau,\bfk)}{a(\tau)}\bigg)' \nnr
&=ka(\tau)\bigg(\frac{\dd\bar{h}_\lambda(\tau,\bfk)}{\dd z} + \frac{1}{z}(1 + \epsilon)\bar{h}_\lambda(\tau,\bfk)\bigg) \,.
\end{align}
These expressions allow us to write the $h_\lambda$ commutator as
\begin{align}
&\big[h_\lambda(\tau,\bfk),p_{\lambda'}(\tau,\bfk')\big] = \big[\bar{h}_\lambda(\tau,\bfk),\bar{p}_{\lambda'}(\tau,\bfk')\big] \nnr
&\qquad= i\big[\bar{h}^R_\lambda(\tau,\bfk),\bar{p}^I_{\lambda'}(\tau,\bfk')\big] + i\big[\bar{h}^I_\lambda(\tau,\bfk),\bar{p}^R_{\lambda'}(\tau,\bfk')\big] \nnr
&\qquad= 2ikz^{\epsilon\, \delta n}\big[\alpha_\lambda(\tau,\bfk),\alpha^\dag_{\lambda'}(\tau,-\bfk')\big]\bigg(\!(1 + 2\rho)\Big(1 + \epsilon\frac{1 + 8\rho}{(1 + \rho)^2}\Big)
+ 4\rho^2z^2\Big(1 - \epsilon\frac{1 + 6\rho}{1 + 2\rho}\Big) + O(z^4,\epsilon^2)\!\bigg) \,, \label{CCR-h-bar}
\end{align}
where
\begin{align}
\delta n = \delta\hat{\gamma} - 1 = \frac{-4\rho}{1 + 2\rho} \,, \label{delta-n}
\end{align}
and $\delta \hat{\gamma}$ is given in (\ref{gammadef}). The $H_\lambda$ commutator also follows in a similar fashion:
\begin{align}
&\big[H_\lambda(\tau,\bfk),P_{\lambda'}(\tau,\bfk')\big] = \big[\bar{H}_\lambda(\tau,\bfk),\bar{P}_{\lambda'}(\tau,\bfk')\big] \nnr
&\qquad= i\big[\bar{H}^R_\lambda(\tau,\bfk),\bar{P}^I_{\lambda'}(\tau,\bfk')\big] + i\big[\bar{H}^I_\lambda(\tau,\bfk),\bar{P}^R_{\lambda'}(\tau,\bfk')\big] \nnr
&\qquad= -2ikz^{\epsilon\,\delta n}\big[\alpha_\lambda(\tau,\bfk),\alpha^\dag_{\lambda'}(\tau,-\bfk')\big]\bigg(4\rho^2z^2\Big(1 - \epsilon\frac{1 + 6\rho}{1 + 2\rho}\Big) + O(z^4,\epsilon^2)\bigg) \,. \label{CCR-H-bar}
\end{align}
We note that (\ref{CCR-h-bar}) and (\ref{CCR-H-bar}) correspond to $O(\epsilon)$ corrections to (\ref{CCR-h}) and (\ref{CCR-H}), as one should expect. As before, we may now combine the expressions above to obtain the commutator for the summed spin-2 modes:
\begin{align}
&\big[h_\lambda(\tau,\bfk) + H_\lambda(\tau,\bfk),p_{\lambda'}(\tau,\bfk') + P_{\lambda'}(\tau,\bfk')\big] \nnr
&\qquad= 2ikz^{\epsilon\,\delta n}\big[\alpha_\lambda(\tau,\bfk),\alpha^\dag_{\lambda'}(\tau,-\bfk')\big](1 + 2\rho)\bigg(1 + \epsilon\frac{1 + 8\rho}{(1 + 2\rho)^2}\bigg) + O(\epsilon^2,z^6) \,. \label{CCR-alpha}
\end{align}
Interestingly, we see that the $\tau$-dependent term (the $z^2$ term in (\ref{CCR-h-bar})) cancels against the same type of term in (\ref{CCR-H-bar}), and that this happens up to and including $O(z^4=k^4\tau^4)$ in (\ref{CCR-alpha}) for the coefficients in (\ref{c3def}) and (\ref{gammadef}).

It is important to note that the last term proportional to $\epsilon$ in (\ref{CCR-alpha}) is $\tau$-independent (its dependence appears at $O(\epsilon^2)$ through $\rho$)
and it is related to the $O(\epsilon)$ normalization of the asymptotic solutions at $O(\epsilon)$, where the normalization at $\epsilon=0$ is fixed. Clearly, the correlation function $\mel{0}{\big(h_\lambda(\tau,\bfk)+H_\lambda(\tau,\bfk)\big)\big(h_{\lambda'}(\tau,\bfk')+H_{\lambda'}(\tau,\bfk')\big)}{0}$ does not depend on 
the change of an overall normalization of the asymptotic solutions, as this can be absorbed by the operator $\alpha_\lambda(\bfk)$. However, a relative change of the normalization between the real and imaginary parts of $f(z)$ does effect this overall normalization. Thus, using only the exact solutions (\ref{fR0}--\ref{gI0}) at $\epsilon=0$ and the asymptotic solutions\footnote{
  The $O(\epsilon)$ correction in General Relativity has been computed in \cite{Stewart:1993bc}.} 
at $O(\epsilon)$, we obviously can not fix the relative normalization. This can only be fixed by the behavior of the solution
for large $z=-k\tau$; the solutions $f(z)$ and $g(z)$ should become proportional to $e^{iz}$ in the large $z$ limit. Therefore, the $\tau$-independent $O(\epsilon)$ term in (\ref{CCR-alpha}) is subject to change.

Though the $\tau$-independent $O(\epsilon)$ term derived above will contribute to the power spectrum, its contribution is less than $O(0.1\%)$, as $\epsilon \lesssim 0.004$. We will thus neglect this small correction in the following discussion\footnote{
  This $O(\epsilon)$ correction will change the amplitude of the tensor power spectrum and so the tensor-to-scalar ratio $r$, which is of $O(\epsilon)$. Therefore, it will be an $O(\epsilon^2)$ contribution  to $r$.}.
Furthermore, according to the discussion at the end of Section \ref{sec:diag-con}, the RHS of (\ref{CCR-alpha}) should be equal to $i\delta_{\lambda\lambda'}\delta^3(\bfk+\bfk')$, since we have taken into account only the massless mode in (\ref{CCR-alpha}), which all together leads to the relation
\begin{align}
\big[\alpha_\lambda(\tau,\bfk),\alpha^\dag_{\lambda'}(\tau,-\bfk')\big] = \frac{(-k\tau)^{-\epsilon\,\delta n}}{2k\big(1 + 2\rho(\tau)\big)}\delta_{\lambda\lambda'}\delta^3(\bfk + \bfk') \,. \label{CCR-alpha-final}
\end{align}
It should be noted that the $\tau$ dependence of the RHS of this expression appears at $O(\epsilon^2)$; the $\tau$ dependence of $(1+2\rho(\tau))^{-1}$ is cancelled by $(-k\tau)^{-\epsilon\,\delta n}$, which follows from
\begin{align}
\frac{\dd}{\dd\tau}(-k\tau)^{\epsilon\,\delta n}\big(1 + 2\rho(\tau)\big) = -k\epsilon\big(4\rho(\tau) + \delta n(1 + 2\rho)\big)(-k\tau)^{\epsilon\,\delta n - 1} + O(\epsilon^2) = O(\epsilon^2) \,,
\end{align}
where we have used (\ref{rho-dash}).

Collecting our results obtained above, we are now able to calculate the power spectrum
\begin{equation}
{\cal P}_{hH}(\bm k, \bm k') = \frac{4}{\Mp^2}\sum_{\lambda=\pm}\mel{0}{\big(h_\lambda(\tau,\bfk) + H_\lambda(\tau,\bfk)\big)\big(h_{\lambda'}(\tau,\bfk') + H_{\lambda'}(\tau,\bfk')\big)}{0} \label{PS-hH-1}
\end{equation}
in the asymptotic limit. To this end, we first note that only $f^R(z)$ among the asymptotic solutions (\ref{ansatz-f}) and (\ref{ansatz-g}) survives as $-\tau$ approaches  zero, so that
\begin{align}
{\cal P}_{hH}(\bm k, \bm k') &\approx \frac{4}{\Mp^2}\sum_{\lambda=\pm}\mel{0}{h^R_\lambda(\tau,\bfk)h^R_{\lambda'}(\tau,\bfk')}{0} = \frac{4}{\Mp^2a^2(\tau)}\sum_{\lambda=\pm}\mel{0}{\bar{h}^R_\lambda(\tau,\bfk)\bar{h}^R_{\lambda'}(\tau,\bfk')}{0} \nnr
&= \frac{8}{\Mp^2a^2(\tau)}(-k\tau)^{2\gamma_1}\mel{0}{\alpha_\lambda(\tau,\bfk)\alpha^\dag_\lambda(\tau,-\bfk')}{0} \nnr
&= \frac{4(-k\tau)^{-2 - \epsilon(2 + \delta n)}}{k\Mp^2a^2(\tau)(1 + 2\rho(\tau))}\delta^3(\bfk + \bfk') \,, \label{PS-hH-2}
\end{align}
where we have used (\ref{vacuum}) and (\ref{CCR-alpha-final}), the definitions of $\gamma_1$ and $\delta n$ are in (\ref{gammadef}) and (\ref{delta-n}), and the relation $\bar{h}^R_\lambda(\tau,\bfk)=(\alpha_\lambda(\tau,\bfk)+\alpha^\dag_\lambda(\tau,-\bfk))f^R(z)$, where $f^R(z)$ is given in (\ref{ansatz-f}).

Crucially, this asymptotic expression is in fact $\tau$-independent, which follows from the following two facts. Firstly, because $\dd a(\tau)/\dd\tau=a(\tau)(a(\tau)\bm{H}(\tau))\approx a(\tau)(1+\epsilon)$, we can write
\begin{align}
a(\tau) \approx a_0\bigg(\frac{-k\tau}{-k\tau_0}\bigg)^{-(1 + \epsilon)} \,, \label{a-tau}
\end{align}
where $a_0=a(\tau_0)$. Secondly, because $\dd\rho(\tau)/\dd\tau\approx2\epsilon\,\rho(\tau)/\tau$, we find
\begin{align}
1 + 2\rho(\tau) &\approx 1 + 2\rho_0\bigg(\frac{-k\tau}{-k\tau_0}\bigg)^{2\epsilon} \approx 1 + 2\rho_0\bigg(1 + 2\epsilon\,\ln\frac{-k\tau}{-k\tau_0}\bigg) \nnr
&= (1 + 2\rho_0)\bigg(1 + \frac{4\epsilon\,\rho_0}{1 + 2\rho_0}\ln\frac{-k\tau}{-k\tau_0}\bigg) \approx (1 + 2\rho_0)\bigg(\frac{-k\tau}{-k\tau_0}\bigg)^{4\epsilon\,\rho_0/(1 + 2\rho_0)} \nnr
&\approx (1 + 2\rho_0)\bigg(\frac{-k\tau}{-k\tau_0}\bigg)^{-\epsilon\,\delta n} \,. \label{1+2rho}
\end{align}
The power spectrum (\ref{PS-hH-2}) can therefore be written as
\begin{align}
{\cal P}_{hH}(\bm k, \bm k') \approx \frac{4(-k\tau_0)^{-2 - \epsilon(2 + \delta n)}}{k\Mp^2a_0^2(1 + 2\rho_0)}\delta^3(\bfk + \bfk') \approx \frac{2\pi^2}{k^3}\Delta^2_t(k)\delta^3(\bfk + \bfk') \,, \label{PS-hH-3}
\end{align}
with
\begin{align}
&\Delta^2_t(k) = A_t(k_*)\bigg(\frac{k}{k_*}\bigg)^{-2\epsilon_*/(1 + 2\delta n_*)} \,,& &A_t(k_*) = \frac{2\bm{H}_*^2}{\pi^2\Mp^2(1 + 2\rho_*)} \,,
\end{align}
where we have identified $\tau_0$ as the horizon-crossing time $\tau_*$ and introduced the pivot scale $k_*=a_*\bm{H}_*\approx-(1+\epsilon_*)\tau_*^{-1}$. We recall that going from (\ref{PS-hH-1}) to (\ref{PS-hH-3}), we have consistently neglected terms whose $\tau$ dependence appears at $O(\epsilon^2)$. 

Finally, the QG tensor-to-scalar ratio in quasi de Sitter is given by
\begin{align}
r = \frac{A_t(k_*)}{A_s(k_*)} = \frac{r_\text{E}}{1 + 2\rho_*} = \frac{r_\text{E}}{1 + 2\bm{H}^2_*/m^2} \,, \label{r-ratio}
\end{align}
where $A_s(k_*)$ is the amplitude of the scalar power spectrum and $r_\text{E}$ stands for the tensor-to-scalar ratio in the absence of the Weyl tensor squared term in the action (\ref{SQG}). The spectral index of the tensor power spectrum\footnote{
  This result is corroborated in \cite{Salvio:2020axm,Salvio:2022mld} where it is derived based on the requirement that the $\delta n$ which appears in \eqref{PS-hH-2} is chosen so that the power spectrum is $\tau$-independent. While this appears to be a logical requirement to make, our derivation based on solving the EOMs independently justifies this requirement, at least to the order we are concerned with.
}, $n_t$, is
\begin{align}
n_t = \frac{\dd\Delta^2_t(k)}{\dd\ln k} = -\epsilon_*(2 + \delta n_*) = -\epsilon_*\frac{2}{1 + 2\rho_*} = \frac{-2\epsilon_*}{1 + 2\bm{H}^2_*/m^2} \,, \label{nt}
\end{align}
which reduces to the General Relativity result, $n_t=-2\epsilon_*$, in the $m\to\infty$ limit \cite{Fabbri:1983us,Abbott:1984fp,Stewart:1993bc}. It is interesting to observe that the correction coming from the Weyl tensor squared term to the tensor-to-scalar ratio (\ref{r-ratio}) and that to the spectral index (\ref{nt}) have the same expression. Additionally, since $r_\text{E}=16\epsilon_*$ in single-field slow-roll models for inflation, (\ref{r-ratio}) and (\ref{nt}) imply an effective change of the slow-roll parameter $\epsilon_*$ to  $\epsilon_*/(1+2 \bm{H}_*/m^2)$.

Some further remarks are in order regarding our result (\ref{nt}). This result agrees with that of \cite{Anselmi:2020lpp}, in which the fakeon prescription is used to eliminate threats to unitarity from the spin-2 ghost, though in their treatment, the massless mode satisfies a second-order DE with a solution proportional to $J_{\nu_t}(-k\tau)\pm iY_{\nu_t}(-k\tau)$, where $\nu_t=3/2+\epsilon_*/(1+2\rho_*)$. Since this solution enters into the tensor power spectrum, it should be compared with
our $ h_\lambda(-k\tau)+H_\lambda(-k\tau)$, however, as we see from the asymptotic solutions (\ref{ansatz-f}--\ref{ansatz-g}), our solution can not be proportional to $J_{\nu_t}(-k\tau)\pm i Y_{\nu_t}(-k\tau)$. Therefore, although the tensor power spectrum (\ref{PS-hH-3}) agrees with the result of \cite{Anselmi:2020lpp} at the lowest nontrivial in the slow-roll approximation, we expect differences to appear at higher orders.

\section{The ghost problem} \label{sec:gprob}

To our knowledge, the ghost problem in curved spacetime is not currently understood to a very significant degree, though it is well-understood in flat spacetime. Thus, in order to proceed, we assume here that the investigations of the ghost problem made for a flat space time can be applied to the present situation. Drawing upon this established knowledge, we will now discuss the matter in two successive epochs; during and after inflation. We also note that we have assumed that the ground state (in the Heisenberg picture) during inflation is the Bunch-Davies vacuum \cite{Bunch:1978yq}, which is an empty vacuum state.

An extraordinary point related to the present discussion is that quantum fluctuations of massless (or nearly massless) modes can become classical field configurations after horizon exit, despite the fact that their vacuum expectation value vanishes (see \cite{Lyth:1984gv,Guth:1985ya,Polarski:1995jg,Lyth:2006qz} and \cite{Martineau:2006ki,Kiefer:2008ku} for discussions in the closed and open system approaches regarding this fact, respectively). These classical configurations are nothing but the seeds of CMB anisotropy and large scale structure of the universe \cite{Mukhanov:1981xt,Hawking:1982cz,Starobinsky:1982ee,Guth:1982ec,Bardeen:1983qw}. For the ghost, since its power spectrum is negative at $\bm k=-\bm k'$ (as we see from (\ref{PS-H0-3})), the power spectrum cannot be understood as an expectation value of the product of two commuting Hermitian fields, $\ssz{H}_{\lambda}(\tau,\bfk)\ssz{H}_{\lambda'}(\tau,\bfk')=\ssz{H}_{\lambda}(\tau,\bfk)\ssz{H}_{\lambda'}^\dag(\tau,-\bfk')$, and thus loses its physical meaning. Crucially however, the ghost mode is massive even in the $m_\text{gh}^2 \to 0$ limit: as we see from (\ref{exact-hH0}), the massive ghost perturbation $\ssz{H}_{\lambda}(\tau,\bfk)$ decreases (decays) like $(-k\tau)^{3/2-\nu}$ as $(-k\tau)\to0$, where $\nu=(1/4-m_\text{gh}^2/{\bm H}^2)^{1/2}\to1/2<3/2$ as $m_\text{gh}\to0$. Consequently, the classicality requirement is not satisfied even if the ghost were an ordinary particle (this includes the case of an imaginary $\nu$) \cite{Starobinsky:1982ee,Lyth:1984gv,Guth:1985ya,Polarski:1995jg,Lyth:2006qz}. Therefore, we may assume that the ghost fluctuation can not become classical i.e.\ Wheeler’s ``decoherence without decoherence'' \cite{Polarski:1995jg} can not occur, and that this is true even beyond the first-order perturbation \cite{Lyth:2006qz}. As long as the ghost fluctuation is quantum mechanical, there is no problem because it does not have any effect on the anisotropy of the universe.

Now, just after the end of inflation, the universe is supposed to undergo reheating, at which time particles are also created, meaning that this is an epoch in which the spin-2 ghost particles may be produced. The maximum  energy (temperature) $E_\text{max}$ in this epoch is assumed to occur just after the end of inflation and is estimated to be $E_\text{max}\sim\big(V_\text{end}^{1/4}T_\text{RH}\big)^{1/2}$, where $V_\text{end}$ is the vacuum energy at the end of inflation and $T_\text{RH}$ is the reheating temperature \cite{Kolb:1990vq,Chung:1998rq}. One should also note that, since $V_\text{end}^{1/4}T_\text{RH}/\Mp^2\ll1$, $E_\text{max}$ can be several orders of magnitude smaller than the Planck scale $\Mp$. This is of central importance, as we establish shortly, because the ghost problem depends strongly on whether the ghost is stable or not, or more precisely, whether asymptotic ghost states exits or not. For completeness, it should also be noted that we would face a completely different situation if the ghost were to be confined like the gluon \cite{Kawasaki:1981gk,Arkani-Hamed:2003pdi,Mukohyama:2009rk}, in which case the calculation of tensor fluctuations in QG should be entirely changed.

Our standpoint here is that the massive ghost particle is fundamental and stable. Though works such as \cite{Donoghue:2019fcb} make claims to the contrary, the in-depth analysis in \cite{Kubo:2023lpz,Kubo:2024ysu} makes it clear that this must be the case. Indeed, one may derive a negative decay rate for the ghost, which does \textit{not} mean that the ghost particle decays, rather, it reflects a mathematical fact that the pole of the ghost propagator is shifted into the (physical) first sheet of complex four momentum squared $p^2$ \cite{Lee:1970iw}. One may use dispersion relations to derive the K\"all\'en-Lehman representation of the ghost propagator and show that the asymptotic ghost states with a pair of conjugate complex masses $m_\text{gh}$ and $m_\text{gh}^*$ survive \cite{Kubo:2024ysu}\footnote{
  In the case of an ordinary unstable particle where the decay width is positive, this does not imply that the pole of the propagator is shifted into the physical first sheet of complex $p^2$, rather, it shifts into the unphysical second sheet of the Riemann surface, meaning that there is no pole at all on the physical sheet. This implies that no asymptotic state exits and that the corresponding unstable state does not contribute to the optical theorem \cite{Veltman:1963th}.}. 
The complex ghost particles can nevertheless be ``produced'' through the scattering of ordinary particles i.e.\ the amplitude for such process does not strictly vanish. This is because, in the presence of complex energy, the conventional Dirac delta function that expresses the energy conservation at each vertex of interaction should be generalized to a complex delta function (a complex distribution) which allows such production without violating energy conservation \cite{Kubo:2023lpz}.

Thus, if spin-two ghost particles are produced during the reheating epoch we encounter a problem with physical unitarity. Although the complex delta function allows ``unsharp'' energy conservation, it also defines a sharp threshold $m_\text{thr}=\Re m_\text{gh}-\Im m_\text{gh}$ (for $\Re m_\text{gh} > \Im m_\text{gh}$), below which ghost particles can not be produced. Therefore, under the kinematic condition that $m_\text{thr}$ is larger than the maximum kinetic energy $E_\text{max}$, physical unitarity is not violated \cite{Kubo:2023lpz} (see also \cite{Kubo:2022jwu}). With this knowledge, we can take something from the physically realistic example of $E_\text{max}$ in \cite{Chung:1998rq}: when $E_\text{max}\sim3\times10^7\,\big(T_\text{RH}/[\text{GeV}]\big)^{1/2}$ GeV, the unitarity requirement $m_\text{gh} \sim m_\text{thr}\lesssim E_\text{max}$ means that $H_*/m_\text{gh}\lesssim4\times10^{6} \big(T_\text{RH}/[\text{GeV}]\big)^{-1/2}$. Interestingly, this means that the correction $2H_*^2/m_\text{gh}^2$ in (\ref{correction-to-r}) can in fact become larger than $O(1)$ for $T_\text{RH}\lesssim10^{12}$ GeV in this case.

\section{Conclusion}

Quadratic gravity is a perturbatively renormalizable theory \cite{Stelle:1976gc} that maintains a quantum-theoretic interpretation as a physically unitary theory, provided that its spin-2 ghost remains a virtual excitation only \cite{Kubo:2022jwu,Kubo:2023lpz}. In other words, when the spin-2 ghost does not appear as a genuine asymptotic state, QG is a sensible quantum field theory that can be treated perturbatively in the usual fashion. In this work, we have studied tensor fluctuations (gravitational waves) in QG around an FLRW background in this regime. By using an auxiliary tensor field, the Weyl tensor squared term in the action, which is both indispensable for renormalizability and the origin of the spin-2 ghost, can be rewritten so as to contain only second-order derivatives \cite{Kubo:2022jwu}. With this trick QG assumes a form in which the theory can be quantized \`a la conventional canonical methods. Using this formalism in pure de Sitter space-time, we have confirmed previously known results on the tensor power spectrum \cite{Clunan:2009er,Myung:2014jha,Myung:2015vya,Salvio:2017xul,Anselmi:2020lpp} in QG.

We have also extended these calculations to quasi de Sitter spacetime where the Hubble parameter varies slowly in time in line with the standard slow-roll consistency condition (\ref{consistency}) \cite{Liddle:1992wi,Turner:1993xz,Copeland:1993jj,Lidsey:1995np}, and found that this condition is indeed satisfied. In the calculation of \cite{Anselmi:2020lpp} in quasi de Sitter space-time, the spin-two ghost is identified as a fakeon \cite{Anselmi:2018kgz,Anselmi:2018tmf} and is eliminated from the asymptotic spectrum via the fakeon prescription, confirming also the slow-roll consistency condition (\ref{consistency}) in the process. Despite this confirmation at the lowest non-trivial order, we have argued that, as the fakeon prescription employs methods that go beyond the conventional wisdom of QFT, certain differences will appear in higher order calculations of the tensor power spectrum. However, to precisely calculate the differences between the conventional QFT and fakeon frameworks, one must solve the system of the coupled DEs (\ref{eom-h}) and (\ref{eom-H}) exactly. Though this task is beyond the scope of the present paper, it will be an interesting problem to tackle in the future.

\clearpage
\section*{Acknowledgments}

We thank Taichiro Kugo for valuable comments and suggestions which have been indispensable for completing the present work. This work was supported in part by the MEXT/JSPS KAKENHI Grant Number 23K03383 (J.\ Kubo).

\bibliography{library,library-kubo}
\bibliographystyle{utphys}

\end{document}